\pdfoutput=1
\documentclass[10pt,twocolumn,aps,
prl,
superscriptaddress, floatfix,notitlepage]{revtex4-1}

\usepackage[utf8]{inputenc}
\usepackage[english]{babel}
\usepackage{graphicx}
\usepackage{amsmath}
\usepackage{amssymb}
\usepackage{bm}
\usepackage[dvipsnames]{xcolor}
\usepackage{bbm}
\usepackage{bbold}

\usepackage[colorlinks, citecolor={blue!50!black}, urlcolor={blue!50!black}]{hyperref}
\usepackage{bookmark}

\graphicspath{{../figures/}}

\newcommand{\pd}{{\phantom\dag}}
\newcommand{\cm}{C_8M}

\begin{document}

\author{D{\'a}niel Varjas}
\email[Electronic address: ]{dvarjas@gmail.com}
\affiliation{QuTech, Delft University of Technology, P.O. Box 4056, 2600 GA Delft, The Netherlands}
\affiliation{Kavli Institute of Nanoscience, Delft University of Technology, P.O. Box 4056, 2600 GA Delft, The Netherlands}
\author{Alexander Lau}
\affiliation{Kavli Institute of Nanoscience, Delft University of Technology, P.O. Box 4056, 2600 GA Delft, The Netherlands}
\author{Kim P{\"o}yh{\"o}nen}
\affiliation{QuTech, Delft University of Technology, P.O. Box 4056, 2600 GA Delft, The Netherlands}
\affiliation{Kavli Institute of Nanoscience, Delft University of Technology, P.O. Box 4056, 2600 GA Delft, The Netherlands}
\author{Anton~R.~Akhmerov}
\affiliation{Kavli Institute of Nanoscience, Delft University of Technology, P.O. Box 4056, 2600 GA Delft, The Netherlands}
\author{Dmitry I. Pikulin}
\affiliation{Microsoft Quantum, Microsoft Station Q, University of California, Santa Barbara, California 93106-6105}
\author{Ion Cosma Fulga}
\affiliation{IFW Dresden and W{\"u}rzburg-Dresden Cluster of Excellence ct.qmat, Helmholtzstr. 20, 01069 Dresden, Germany}

\title{Topological phases without crystalline counterparts}

\begin{abstract}
Recent years saw the complete classification of topological band structures, revealing an abundance of topological crystalline insulators.
Here we theoretically demonstrate the existence of topological materials beyond this framework, protected by quasicrystalline symmetries.
We construct a higher-order topological phase protected by a point group symmetry that is impossible in any crystalline system.
Our tight-binding model describes a superconductor on a quasicrystalline Ammann–Beenker tiling which hosts localized Majorana zero modes at the corners of an octagonal sample.
The Majorana modes are protected by particle-hole symmetry and by the combination of an 8-fold rotation and in-plane reflection symmetry.
We find a bulk topological invariant associated with the presence of these zero modes, and show that they are robust against large symmetry preserving deformations, as long as the bulk remains gapped.
The nontrivial bulk topology of this phase falls outside all currently known classification schemes.
\end{abstract}
\maketitle


\emph{Introduction.} ---
All topological phases known to date can exist in crystalline systems.
Strong topological insulators (TI) occur in crystals~\cite{hazan2010toporeview,qi2011topological}, in quasicrystals~\cite{shechtman1984metallic,baake2013aperiodic,zoorob2000complete,steurer2007photonic,kraus2012topological,
verbin2013topoquasicrystal,fulga2016aperiodic}, and even in amorphous systems~\cite{fulga2014statistical,agarwala2017topoamorphous,kim2018amorphous}, and show gapless modes on any boundary, as a consequence of the topologically nontrivial gapped bulk.
Weak topological insulators and topological crystalline insulators, on the other hand, rely on crystal symmetries~\cite{fu2007topological,rasche2013wti, fu2011topological, hughes2011inversion, chiu2016classification}.
Their gapless topological states appear only on boundaries preserving, at least on average, a subset of lattice symmetries~\cite{ringel2012strong,lau2016mirror}.

In contrast, in higher-order topological insulators (HOTI) both the bulk and the boundaries are gapped.
Instead, the protected gapless modes form at the intersections of two or more boundaries---the corners and hinges of a crystal~\cite{benalcazar2017science, benalcazar2017prb,schindler2018hoti,trifunovic2019hoti,langbehn2017mirrorhoti, geier2018hoti,franca2018hoti}.
Unlike in topological crystalline insulators, the corners/hinges may break the lattice symmetry responsible for protecting the HOTI.
In those cases, the protection of the boundary modes relies on a discrete symmetry of the entire finite-sized sample.
Examples of HOTIs enabled by global symmetries include a three-dimensional (3D) TI placed in a magnetic field~\cite{sitte2012hinges}, hosting chiral hinge modes protected by inversion symmetry, as well as elemental bismuth~\cite{schindler2018bismuth}, with helical hinge modes protected by time-reversal symmetry (TRS), 3-fold rotation, and inversion.

Since the set of allowed crystal symmetries is known, it is possible to list all weak, crystalline, and higher-order topological insulators that appear in a crystal.
This program has been carried out throughout the past decade, starting with the effect of single symmetries such as mirror or inversion, followed by considering the effect of any order-two symmetry~\cite{chiu2013mirror,shiozaki2014ordertwo}.
Today, the topological classification spans all known non-magnetic crystalline compounds ~\cite{kruthoff2017classification,po2017classification, Bradlyn2017}.
Furthermore, the possible band topologies of free fermions have been listed for all 528 two-dimensional (2D) and 1651 3D magnetic space groups~\cite{watanabe2018classification}.

\begin{figure}[tb]
    \includegraphics[width=\columnwidth]{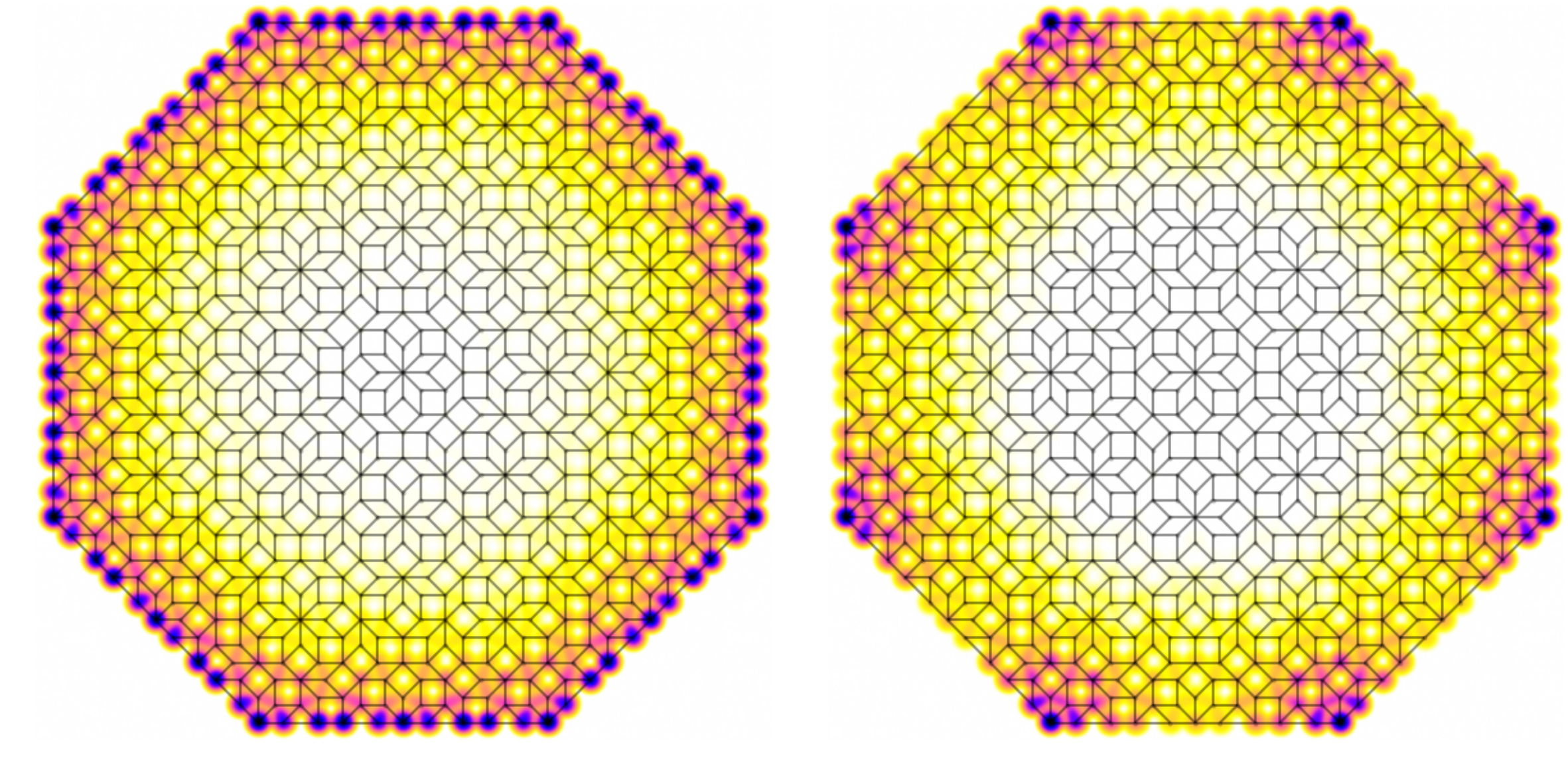}
    \caption{We define a tight-binding model on an 8-fold symmetric patch of the Ammann-Beenker tiling by associating a site to each vertex and a hopping to each of the edges that connect neighboring vertexes.
Both panels show the real-space distributions of the wavefunction amplitudes in the eight lowest energy states of the model defined in Eqs.~\eqref{eq:Hreal} and~\eqref{eq:V8} for $\Delta=t=1$ and $\mu=-1.7$.
Darker colors denote larger amplitudes.
For $V=0$ (left), the system hosts counter-propagating Majorana modes on any edge, protected by mirror symmetry.
Setting $V=1$ (right) gaps out the edge, leading to a HOTI phase.
A single Majorana zero mode is localized to each of the eight corners of the tiling.}
    \label{fig:AB_tiling}
\end{figure}

In this work, we extend the existing classification of topological phases by constructing a HOTI phase that is incompatible with a crystalline symmetry, and was therefore overlooked in the previous works.
This topological phase relies on the combination of an 8-fold rotation and an in-plane reflection.
Its hallmark signature is the presence of eight Majorana zero modes bound to the corners of a finite-sized octagonal sample.
These modes are robust against any symmetry-preserving perturbation, provided the bulk remains gapped.
Because 8-fold rotations are forbidden in two dimensions by the crystallographic restriction theorem, the resulting phase has no crystalline counterpart.
We propose a modified notion of symmetry protection of HOTI phases applicable to locally generated quasiperiodic Hamiltonians.
Using this, we show that the protection of the corner modes does not rely on global symmetry of the sample.
We pin down the nontrivial nature of this phase by studying zero modes at topological defects and identifying a bulk topological invariant that determines the formation of Majorana corner modes.


\emph{Model.} ---
Our starting point is a tight-binding model describing a pair of oppositely spin-polarized $p \pm ip$ topological superconductors in class D~\cite{read2000pwave}.
The real-space Bogoliubov-de Gennes Hamiltonian is obtained by associating sites and hoppings to the vertexes and edges of an 8-fold symmetric Ammann–Beenker (AB) tiling~\cite{grunbaum1986abtiling,beenker1982abtiling} (see Fig.~\ref{fig:AB_tiling}):
\begin{equation}\label{eq:Hreal}
 {\cal H} = \sum_j {\boldsymbol \Psi}^\dag_j {\cal H}^\pd_j {\boldsymbol \Psi}^\pd_j + \sum_{\langle j, k \rangle} {\boldsymbol \Psi}^\dag_j {\cal H}^\pd_{jk} {\boldsymbol \Psi}^\pd_k,
\end{equation}
with ${\boldsymbol \Psi}^\dag_j = (\psi^\dag_{j, \uparrow}, \psi^\pd_{j, \uparrow}, \psi^\dag_{j, \downarrow}, \psi^\pd_{j, \downarrow})$, $\psi^{\dag}_{j, \sigma}$ the fermionic creation operator for a particle on site $j$ with spin $\sigma$, and with $\langle\ldots\rangle$ denoting sites connected by a bond (see Fig.~\ref{fig:AB_tiling}). The onsite Hamiltonian is
\begin{equation}\label{eq:Hons}
 {\cal H}_j = \mu \sigma_z\tau_z ,
\end{equation}
where $\mu$ is the chemical potential, and Pauli matrices $\tau$ and $\sigma$ act on the electron-hole and spin degrees of freedom, respectively.
The hopping terms have the form
\begin{equation}\label{eq:Hhop}
 {\cal H}_{jk} = \frac{t}{2} \sigma_z \tau_z+\frac{\Delta}{2i} \left[ \cos(\alpha_{jk})\sigma_z\tau_x+\sin(\alpha_{jk})\sigma_z\tau_y \right],
\end{equation}
where $t$ is the normal hopping strength, $\Delta$ the p-wave pairing strength, and $\alpha_{jk}$ is the angle formed by the hopping with respect to the horizontal direction.

The system obeys particle-hole symmetry (PHS), $\{ {\cal H}, {\cal P} \}=0$, with an anti-unitary operator ${\cal P} = \tau_x \sigma_0 \mathbb{1} {\cal K}$, where ${\cal K}$ denotes complex conjugation and $\mathbb{1}$ is the identity operator in the space spanned by the sites of the tiling.
In addition, Eq.~\eqref{eq:Hreal} has an in-plane mirror symmetry, $[ {\cal H}, M ]=0$, with $M = \tau_0\sigma_z\mathbb{1}$.
Moreover, due to the shape of the tiling, the finite-sized model obeys a global 8-fold rotation symmetry about its center, $[{\cal H},C_8] =0$.
The rotation operator is
\begin{equation}\label{eq:C8}
 C_8 = \exp \left(-i \frac{\pi}{8} \sigma_0\tau_z \right) {\cal R},
\end{equation}
where ${\cal R}$ is an orthogonal matrix permuting the sites of the tiling to rotate the whole system by an angle of $\pi/4$.

For $t=\Delta=1$ and $\mu=-1.7$, the model describes a bilayer system of two 2D class D topological superconductors with opposite Chern numbers, hosting a pair of counter-propagating Majorana edge modes on its boundary (see Fig.~\ref{fig:AB_tiling}).
The edge modes are prevented from gapping out by the in-plane reflection symmetry.
To obtain a HOTI, we introduce a perturbation that breaks both the reflection and rotation symmetries, but preserves their product $\cm$.
We modify the hoppings by adding the term
\begin{equation}\label{eq:V8}
{\cal V}  =  \sum_{\langle j, k \rangle} {\boldsymbol \Psi}^\dag_j {\cal V}^\pd_{jk} {\boldsymbol \Psi}^\pd_k, \quad
{\cal V}_{jk} = \frac{V}{2} \sigma_y \tau_0 \cos\left( 4\alpha_{jk} \right).
\end{equation}
It anti-commutes with the reflection symmetry, $\{ {\cal V}, M\}=0$, and opens a gap in the edge spectrum.
However, it also anti-commutes with the 8-fold rotation, such that the gap of the edge states changes sign a total of eight times across the perimeter of the system.
This results in the formation of eight Majorana zero modes, as shown in Fig.~\ref{fig:AB_tiling}.
These modes are localized at the corners of the octagonal sample and are separated from all other states by an energy gap.


\emph{Protected corner modes.} ---
Majorana zero modes bound to the corners of the octagonal tiling are a manifestation of the nontrivial bulk topology of the HOTI.
As long as the tiling obeys PHS and the global $\cm$ constraint, the gapless corner states cannot be removed by any perturbation restricted to the system boundary.
There is an intuitive explanation for this (see also Ref.~\cite{schindler2018hoti}): the minimal surface manipulation compatible with PHS and $\cm$ consists of gluing a Kitaev chain onto each of the eight edges of the tiling, such that adjacent chains are mapped onto each other under $\cm$.
This process changes the number of corner Majoranas by an even number and the original zero modes cannot gap out. This suggests that the octagonal HOTI has a $\mathbb{Z}_2$ classification.

\begin{figure}[tb]
	\includegraphics[width=\columnwidth]{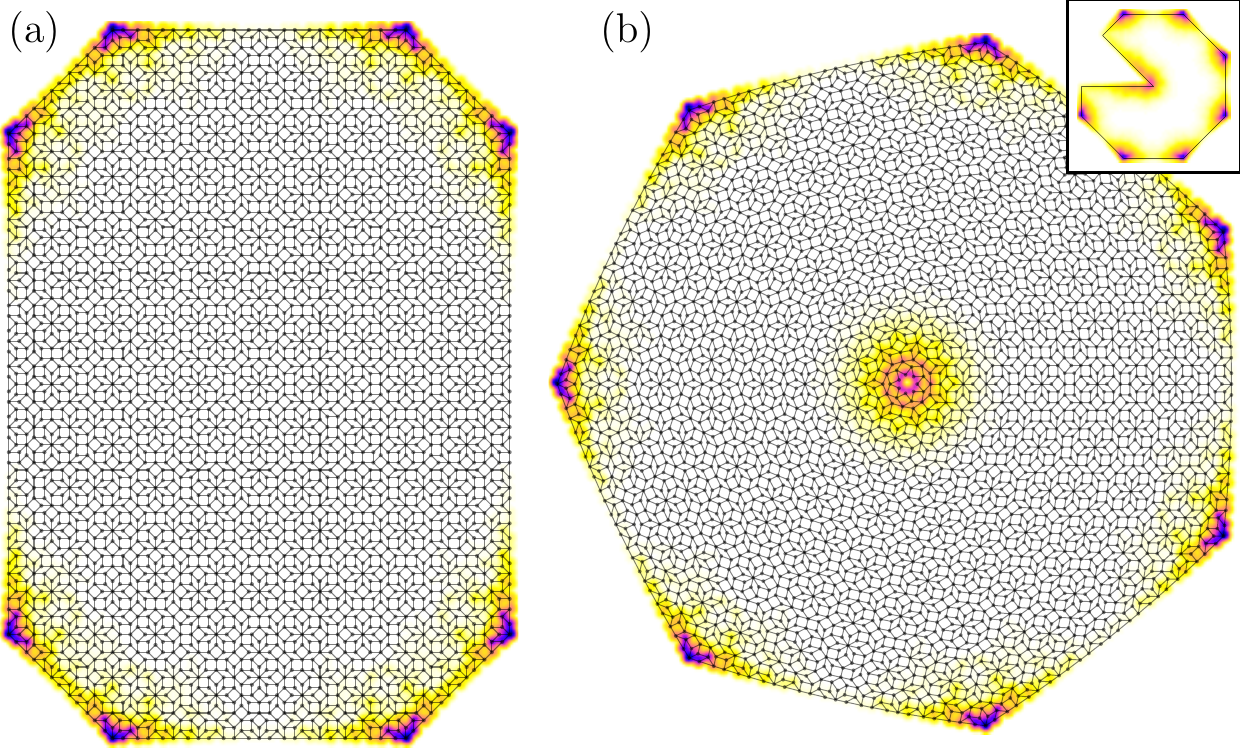}
    \caption{Wave function amplitudes of 8 zero modes in various finite geometries.
    a) Asymmetric sample with corners locally identical to corners of a symmetric sample.
    b) Sample with a $\cm$ defect.
    Away from the defect at the center, the system is locally identical to the original model.
    Inset: Sample with one octant cut out, but without gluing together the two sides of the cut.
    }
    \label{fig:disclination}
\end{figure}

To verify that the corner states are not merely an artifact of an exact $\cm$ symmetry of the entire sample, we also consider asymmetric cutouts of the quasicrystal.
The quasiperiodicity of a quasicrystal implies that any finite region of an infinite sample repeats infinitely many times~\cite{baake2013aperiodic}.
Hence, there are infinitely many locations in the quasicrystal that look identical to the vicinity of a corner of an exactly 8-fold symmetric sample at a scale much larger than the extent of the bound state.
By the locality of the Hamiltonian, such a corner (in either a semi-infinite system or an asymmetric finite sample) will also host a Majorana zero mode, as illustrated in Fig.~\ref{fig:disclination}(a).
These zero modes are ``extrinsic''~\cite{geier2018hoti}, as there is no exact symmetry relating the two edges emanating from such a corner.
This implies that, analogous to crystalline HOTIs, attaching a Kitaev wire to one of the edges but not to the other does not break any symmetries and the zero mode can, in principle, be gapped out by an edge perturbation.

Therefore, we impose the physical restriction of \emph{quasiperiodicity} on the Hamiltonians we consider in the following.
We demand that the Hamiltonian is generated by a local, 8-fold symmetric rule: every term is determined by the quasicrystal configuration in a finite radius environment, in a symmetric fashion.
The quasiperiodicity of the tiling means that the semi-infinite edges emanating from an approximately symmetric corner, while not exact symmetry images, are indistinguishable by only inspecting finite regions.
This prevents a deformation of the Hamiltonian that produces a gapped Kitaev chain on only one of these edges, resulting in protected corner modes.


\emph{Disclination modes} ---
We now prove that the phase discussed above is indeed a \emph{bulk} topological phase protected by $\cm$ symmetry by showing that point-like fluxes (topological defects) of this symmetry capture a Majorana zero mode~\cite{hughes2013disclinations,hughes2014disclinations}.
The $\cm$ flux is inserted into the system by the following procedure (similar to Ref.~\cite{geier2019disclinations}):
first, we take a large 8-fold symmetric sample and cut out one octant bordered by a cut $\mathcal{C}$, connecting the center of the tiling with the boundary, and by its symmetry image $C_8 \mathcal{C}$ (see inset of Fig.~\ref{fig:disclination}(b)).
Then, we glue the two sides of the cut back together by identifying sites on the two sides of the cut related by $\cm$ symmetry.
The hoppings across the cut are $\mathcal{H}_{C_8 j,k} = U_{\cm} \mathcal{H}_{j, k}$, where $C_8 j$ is the $C_8$ image of site $j$ and $U_{\cm}$ is the onsite unitary action of the $\cm$ symmetry.
The cut $\mathcal{C}$, similar to a Dirac string, is not detectable locally.
Indeed, applying a basis transformation $U_{\cm}$ to a single site neighboring the cut (and identity elsewhere) moves the site to the other side of the cut.
This makes the location of the cut basis dependent and locally indistinguishable from the bulk with no cut, with the exception of the center of the system where the cut terminates.

As illustrated in Fig.~\ref{fig:disclination}(b), the resulting sample has eight Majorana zero modes: seven at the corners and one at the disclination core.
The disclination mode cannot be removed without closing the bulk gap, proving that the HOTI phase is separated from the trivial phase by a bulk phase transition.

\emph{Bulk topological invariant.} ---
We now develop an invariant that characterizes the bulk topology of the quasicrystalline system.
For this purpose, we consider the momentum-dependent effective Hamiltonian $H_{\rm eff} = G_{\rm eff}^{-1}$, defined through the projection of the single-particle Green's function onto plane-wave states:
\begin{equation}
G_{\rm eff} (k)_{n,m}  = \left\langle k, n \right| G \left| k, m \right\rangle,
\end{equation}
where $\left| k, n \right\rangle$ is a normalized plane-wave state with nonzero amplitude only in the local orbital $n$, and $G = \lim_{\eta\to0} \left( H + i \eta\right)^{-1}$ is the zero-energy Green's function of the full Hamiltonian (see Supplemental Material).

An important property of $H_{\rm eff}$ is that its gap closes only when the gap of the full Hamiltonian closes.
We are going to use this to construct topological invariants: if an invariant defined in terms of $H_{\rm eff}$ can only change when the gap in $H_{\rm eff}$ closes, it implies a bulk phase transition of the full Hamiltonian.
The classification we derive below is thus a subset of the full topological classification of $\cm$ symmetric systems.

To define the topological invariant we inspect the symmetry representations of $\cm$ and $\mathcal{P}$ acting on $H_{\rm eff}$ at the $C_8$-invariant momentum $k=0$~\cite{hughes2013disclinations}.
The eigenvalues of $\cm$ have the form $\omega_n = \exp\left( i \frac{\pi}{8} n \right)$, with $n = [\pm 1, \pm 3, \pm 5, \pm 7]$, and eigenstates $\left| n\right\rangle$ and $\left| -n \right\rangle$ are related by $\mathcal{P}$.
By restricting $H_{\rm eff}(k=0)$ to $\cm$ eigensubspaces of $\omega_{\pm n}$, we calculate the zero-dimensional class D invariant of each block, which is the sign of the Pfaffian in the Majorana basis.
This yields $\nu_{n, k=0} = \pm 1$, for $n \in [1, 3, 5, 7]$, resulting in a $\mathbb{Z}_2^4$ classification.
In our model, $H_{\rm eff}(k=0)$ has two invariant blocks corresponding to pairs of $n=\pm 1$ and $\pm 7$, respectively, while there are no states in the local Hilbert space corresponding to the other $\cm$ eigenstates with $n=\pm 3, \pm 5$.
We find that $H_{\rm eff}(k=0)$ goes through a band inversion at $\mu \approx -2$ when both Pfaffians switch sign. This, however, cannot be a stable topological invariant, as it also distinguishes different atomic insulators with on-site Hamiltonians of opposite sign and vanishing hoppings.

To provide an invariant that is insensitive to addition of atomic insulators, we invoke the cut-and-project method generating the 2D AB tiling from a 4-dimensional (4D) cubic lattice (see Ref.~\onlinecite{baake2013aperiodic} and Supplemental Material).
Plane-wave states in the 4D Brillouin zone provide a complete basis for all states on the 4D lattice, and an overcomplete basis for the quasicrystal.
Some of these plane waves cannot be exactly represented by purely 2D plane waves, but can be approximated by those to arbitrary precision.
We call these patterns of complex phases on the quasicrystal \emph{generalized plane waves}.
The generalized plane waves important for the topological invariant are the ones at 4D momenta invariant under $C_8$ modulo reciprocal lattice vectors.
Those are $\Gamma=(0,0,0,0)\equiv 0$, which we have already discussed above, and $\Pi = (\pi, \pi, \pi, \pi)$.
The latter produces alternating $\pm$ signs on nearest-neighbor sites of the quasicrystal.
Looking at the symmetry representation of $H_{\rm eff} (k=\Pi)$, we find a band inversion at $\mu \approx 2$, which is similar to the one of $H_{\rm eff} (k=0)$ at $\mu \approx -2$.
As a consequence, $\nu_{n, 0} = -\nu_{n, \Pi}$ for $n \in [1, 7]$ in the range $-2 \lesssim \mu \lesssim 2$.
Stable $\mathbb{Z}_2$ invariants are therefore defined by $\nu_n = \nu_{n, 0} / \nu_{n, \Pi}$.

\begin{figure}[tb]
    \includegraphics[width=\columnwidth]{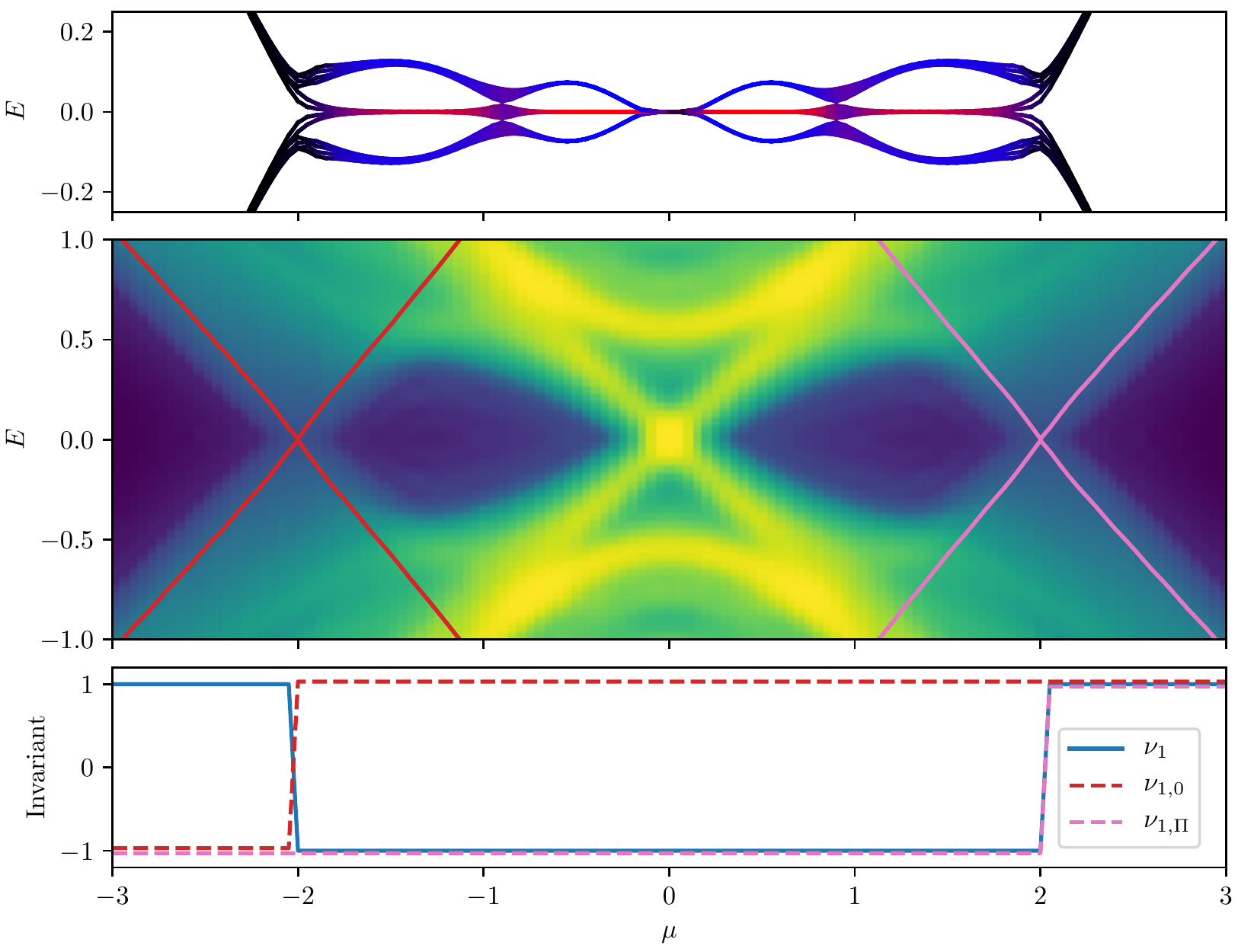}
    \caption{Topological phase transitions in the quasicrystal HOTI model as a function of chemical potential $\mu$ with $t = 1$, $\Delta = 2$ and $V = 1.5$.
    Top panel: spectrum of the 24 states closest to zero energy in a finite sample.
    The line color shows the weight of the state on the corners (red), edges (blue) and bulk (black).
    The bulk gap closes at $\mu\approx \pm 2$, delimiting the phase with eight Majorana corner modes.
    The edge gap closes at $\mu \approx \pm 0.9$ and the bulk gap closes around $\mu=0$ without affecting the topological properties.
    Middle panel: evolution of the bulk density of states, with lighter colors denoting larger densities.
    Overlaid is the spectrum of the effective Hamiltonian at $k=0$ (red) and $k=\Pi$ (pink).
    Bottom panel: topological invariants $\nu_{1, 0}$, $\nu_{1, \Pi}$ and $\nu_1 = \nu_{1, 0} / \nu_{1, \Pi}$.
    }
    \label{fig:invariant}
\end{figure}

In the atomic limit, we have $\nu_n = 1$.
Thus, the nontrivial value signals an obstructed atomic limit.
Moreover, we find that phases with both gapped bulk and gapped edges have only two independent invariants, since $\nu_1 = \nu_7$ and $\nu_3 = \nu_5$ (see Supplemental Material).
In the topological phase, our model has $\nu_1 = \nu_7 = -1$ and $\nu_3 = \nu_5 = +1$, as illustrated in Fig.~\ref{fig:invariant}.
The $\mathbb{Z}_2$ invariant characterizing the presence of corner Majoranas is the product $\nu_1 \nu_3$, while the corner modes do not distinguish between other phases in the richer $\mathbb{Z}_2^2$ bulk classification.


\emph{Discussion.} ---
We have introduced a new type of topological phase that does not fit into any existing classification.
The topological protection of this quasicrystalline HOTI explicitly requires broken translation symmetry, since it is protected by a point group symmetry incompatible with any periodic crystal structure in two or three dimensions.
In the nontrivial phase, both the bulk and the edges are gapped, whereas eight Majorana zero modes are bound to the corners of the octagonal tiling.
These modes are associated with a nontrivial bulk invariant and are robust against symmetry-preserving perturbations that do not close the bulk gap.

While we have treated the special case of a class D topological superconductor with $\cm$ symmetry, the ideas we have presented generalize to a wider range of systems.
It should be possible to extend our work to other symmetry classes, other point group symmetries, and higher dimensions.
We note, however, that the basic line of argument we used to construct the model, reliant on an alternating sign of the mass term at the boundary, does not work for odd rotations, e.g. $C_5$.
For this, it would be necessary to introduce topological protection in another manner.

Our investigation opens several directions for future work.
First, while we have shown a single example as a proof of principle, the range of possible, purely aperiodic topological insulators and semimetals remains unknown.
Furthermore, it is also unclear which tools would be required to classify such systems, given that translation symmetry must be broken, whereas most existing approaches explicitly rely on momentum space.
One might consider, instead, real-space topological invariants, including the ones defined for finite systems with boundaries, as done in Ref.~\onlinecite{loring2015ktheory} for strong topological insulators.
An interesting direction to explore would be to consider classes of quasicrystals obtained by a cut-and-project method from a higher-dimensional periodic lattice, such as the one we used here, and attempt a topological classification via dimensional reduction.
The results of this approach will, however, be limited, since there are quasicrystals not obtainable by such a method.
Lastly, the new methods explored here are applicable to crystalline HOTI's as well.
To show that we found a bulk topological phase, we introduced the notion of a quasiperiodic Hamiltonian, where terms are only sensitive to the quasicrystal configuration in a finite radius environment.
This notion of locality also applies to crystalline, disordered, and amorphous materials, promising a new direction to establish the topological protection of ``extrinsic" corner modes via bulk invariants.

Finally, there is the question of how such a topological phase may be observed experimentally.
While we can predict that this $C_8M$ protected phase will never be realized in any crystalline system, it may be possible to obtain 8-fold symmetry protected corner modes in the recently discovered superconducting quasicrystals \cite{kamiya2018quasisc,araujo2019quasisc}.
Alternatively, one may consider a variety of so-called ``topological simulators'',
including ultracold atoms~\cite{bloch2012simulations,viebahn2019quasicold,corcovilos2019quasicold}, photonic crystals~\cite{lu2014topophoto,ozawa2019topophoto}, coupled electronic circuit elements (called topolectric circuits~\cite{lee2018topolectric}), as well as acoustic and mechanical meta-materials~\cite{susstrunk2015springs,nash2015topomechanic}.
These systems allow for a site connectivity bypassing the chemical constraints inherent in crystal growth processes, and have been successfully used to demonstrate both higher-order topological phases~\cite{Noh2018, xie2018hoti,xue2018hoti}, as well as topologically nontrivial quasicrystals~\cite{verbin2013quasiphotonic,bandres2016quasiphotonic}.

\emph{Author contributions.} ---
I. C. Fulga constructed the model used in the manuscript, initiated, and oversaw the project.
All authors took part in an extensive survey of topological invariants to characterize the system.
D. Varjas conceived and carried out the analysis based on topological defects and bulk topological invariants.
I. C. Fulga and D. Varjas performed the numerical calculations, D. Varjas and A. Akhmerov produced the figures in the manuscript.
All authors took part in formulating the results and writing the manuscript.

\emph{Acknowledgments.} ---
We thank Ulrike Nietsche for technical assistance.
We are grateful to P. Perez-Piskunow for helpful discussions about the kernel polynomial method (KPM) and the use of his KPM Green's function code.
We thank P. Perez-Piskunow and M. Fruchart for sharing the KPM-based method to calculate the Chern number in disordered systems~\cite{Varjas2019}.
This work was supported by the DFG through the W{\"u}rzburg-Dresden Cluster of Excellence on Complexity and Topology in Quantum Matter -- \textit{ct.qmat} (EXC 2147, project-id 39085490).
This work was supported by ERC Starting Grant 638760, NWO VIDI grant 680-47-53, the US Office of Naval Research, and through the subsidy for top consortia for knowledge and innovation (TKl toeslag) by the Dutch ministry of economic affairs.

\bibliography{bibliography}

\begin{thebibliography}{10}
\bibitem{overcomplete}
Note that the ordinary plane-wave basis in 2D is also overcomplete, as it is a
  complete basis for continuum states on the plane that are defined everywhere,
  not only at the sites of the QC.
\end{thebibliography}

\clearpage

\onecolumngrid

\appendix

\section{\large{Supplemental Material to: Topological phases without crystalline counterparts}}
In this Supplemental Material we describe the algorithm used to construct the tiling, as well as the effect of different edge terminations. Further, we provide details on the symmetries and spectrum of the system, as well as its topological invariant.

\section{Tiling construction}

The Ammann-Beenker tiling~\cite{beenker1982abtiling,grunbaum1986abtiling} consists of two primitive tiles: a rhombus, with the small angle equal to $\pi/4$, and a square.
To construct large patches of this quasicrystal (QC), we use an iterative subdivision procedure [summarized in Fig.~\ref{fig:tiling_construction}(a)] applied to an initially small tiling. The construction involves three steps:

\begin{figure}[h]
    \includegraphics[width=\columnwidth]{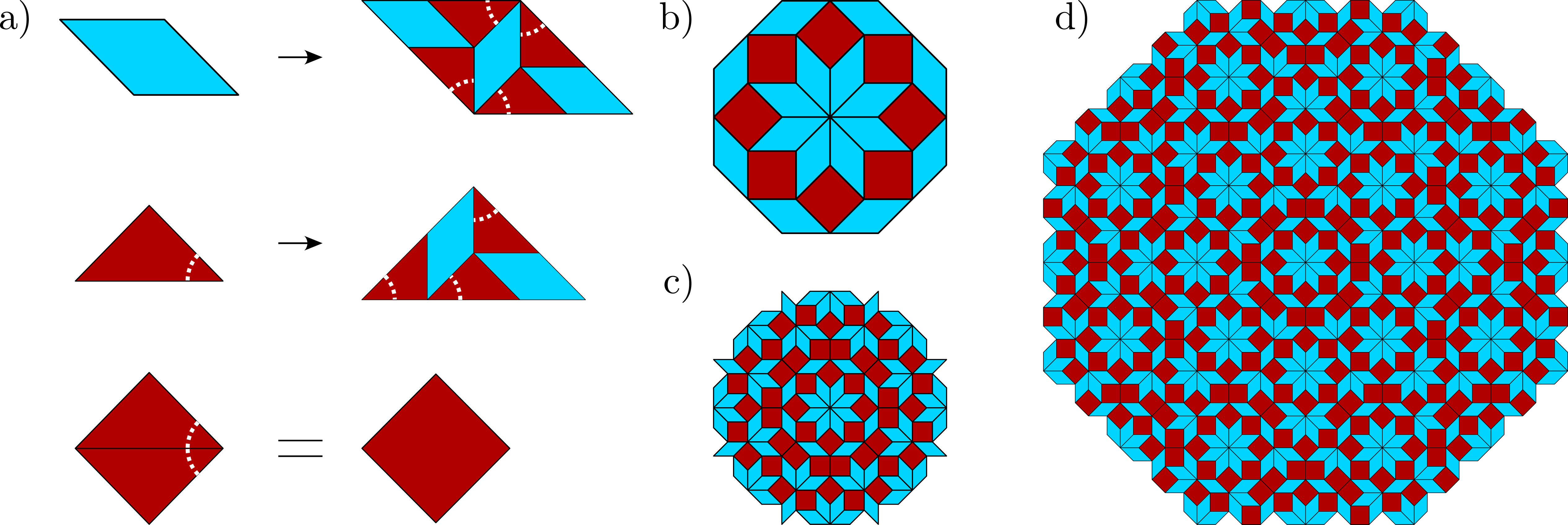}
    \caption{(a) Tile replacement rules used to generate large quasicrystal patches.
    (b) Initial tiling used in the subdivision process.
    (c),~(d) Tilings obtained after $N=1$ and $N=2$ subdivisions, respectively.}
    \label{fig:tiling_construction}
\end{figure}

\begin{enumerate}
 \item Starting from the tiling of Fig.~\ref{fig:tiling_construction}(b), convert all squares into decorated triangles, as indicated in Fig.~\ref{fig:tiling_construction}(a).
 \item Apply the subdivision rules of Fig.~\ref{fig:tiling_construction}(a) $N$ times.
 \item Convert all pairs of triangles back into squares and remove the unpaired edge triangles.
\end{enumerate}

This procedure guarantees that an initially 8-fold symmetric tiling will remain 8-fold symmetric after any number of iterations.
Examples of tilings obtained after $N=1$ and $N=2$ applications of the subdivision process to the initial tiling from Fig.~\ref{fig:tiling_construction}(b) are shown in Fig.~\ref{fig:tiling_construction}(c) and~(d), respectively.

\section{Cut-and-project method and generalized plane waves}

An alternate way to construct the AB tiling is by the cut-and-project method \cite{baake2013aperiodic}.
We start from a four-dimensional (4D) cubic lattice that has the 8-fold symmetry
\begin{equation}
C_8 = \begin{pmatrix}
             0 & 0 & 0 & -1 \\
             1 & 0 & 0 & 0 \\
             0 & -1 & 0 & 0 \\
             0 & 0 & -1 & 0 \\
            \end{pmatrix}.
\end{equation}
This symmetry has two two-dimensional (2D) invariant subspaces, termed parallel and perpendicular, spanned by the row vectors
\begin{eqnarray}
V_{\parallel} &=& \frac{1}{2}\begin{pmatrix}
             -1 & 0 & -1 & \sqrt{2} \\
             1 & \sqrt{2} & -1 & 0
            \end{pmatrix}, \\
V_{\perp} &=& \frac{1}{2}\begin{pmatrix}
             -1 & 0 & -1 & -\sqrt{2} \\
             -1 & \sqrt{2} & 1 & 0
            \end{pmatrix}.
\end{eqnarray}
In both of these subspaces, $C_8$ is represented as an 8-fold rotation in 2D.
To construct the quasicrystal, we remove all sites from the 4D lattice that do not fall within the octagon in the perpendicular space that is the projection of the 4D unit cell centered at the origin.
The coordinates of the remaining sites are then projected onto the parallel space.
The information about the perpendicular position of the atoms is not fully lost:
The perpendicular position is in one-to-one correspondence with the local environment of the site.
The larger the environment observed, the more precisely the perpendicular position can be determined.
Moreover, the nearest-neighbor bonds are projections of bonds in the 4D crystal, where the projection uniquely determines the original bond.
This means that the relative position in parallel space determines the relative position in perpendicular space.

Our goal is to find a basis in the Hilbert space of the tight-binding model corresponding to the infinite quasicrystal where the quasiperiodicity and the homogeneity of the system at large scales is manifest.
We start with the plane-wave basis on the 4D parent crystal, for which plane waves with wave vectors in the first Brillouin zone form an orthogonal basis.
We call states obtained by the restriction of 4D plane waves to the QC \emph{generalized plane waves}.
Besides ordinary 2D plane waves, these include other patterns of complex phases where the phase difference between nearby sites is locally determined.
This is exemplified by the $\Pi = (\pi, \pi, \pi, \pi)$ state used in the main text.
The phase difference in this pattern follows a simple local rule: $-1$ for every nearest-neighbor link.
In the regular plane-wave basis, however, this pattern is an infinite linear combination of plane waves.
It is, hence, useful to use these states to describe states on the quasicrystal.

As the QC is a subspace of the 4D-crystal Hilbert space, the generalized plane-wave basis is overcomplete~\cite{overcomplete}.
To illustrate this, consider two 4D $k$-vectors that have the same projection onto the parallel space: $k = k_{\parallel} + k_{\perp}$, $k' = k_{\parallel} + k'_{\perp}$.
The inner product of the two plane waves restricted to the QC is
\begin{equation}
 \left\langle k \middle| k' \right\rangle = \sum_{R\in QC} e^{i R \left(k'_{\perp} - k_{\perp} \right)} \propto \int d^2 r_{\perp} e^{i r_{\perp} \left(k'_{\perp} - k_{\perp} \right)},
\end{equation}
where the sum runs over all sites in the QC.
In the second step, we use that the perpendicular coordinates of the sites in the QC are uniformly distributed in the octagonal cross section and the sum can be replaced by an integral over the octagon in the perpendicular space.
This equation provides the condition for generalized plane waves with the same parallel component to be orthogonal.

Another important property is that generalized plane waves differing by a 4D reciprocal lattice vector are identical.
As a consequence, ordinary plane waves ($k_{\perp} = 0$) that differ in the projection of a 4D reciprocal lattice vector $G$ are typically not orthogonal, as can be seen using the previous equation.
This is the reason for the Bragg peaks observed in quasicrystals appearing at integer linear combinations of a set of $G_{\parallel}$-vectors, the projections of the 4D reciprocal lattice vectors onto the parallel space.
As these projections are irrationally related, the integer linear combinations form a dense subset of $\mathbb{R}^2$, but most of them have very small overlap with the $k=0$ state.

\section{Effect of edge termination}

\begin{figure}[h]
    \mbox{\includegraphics[width=0.5\columnwidth]{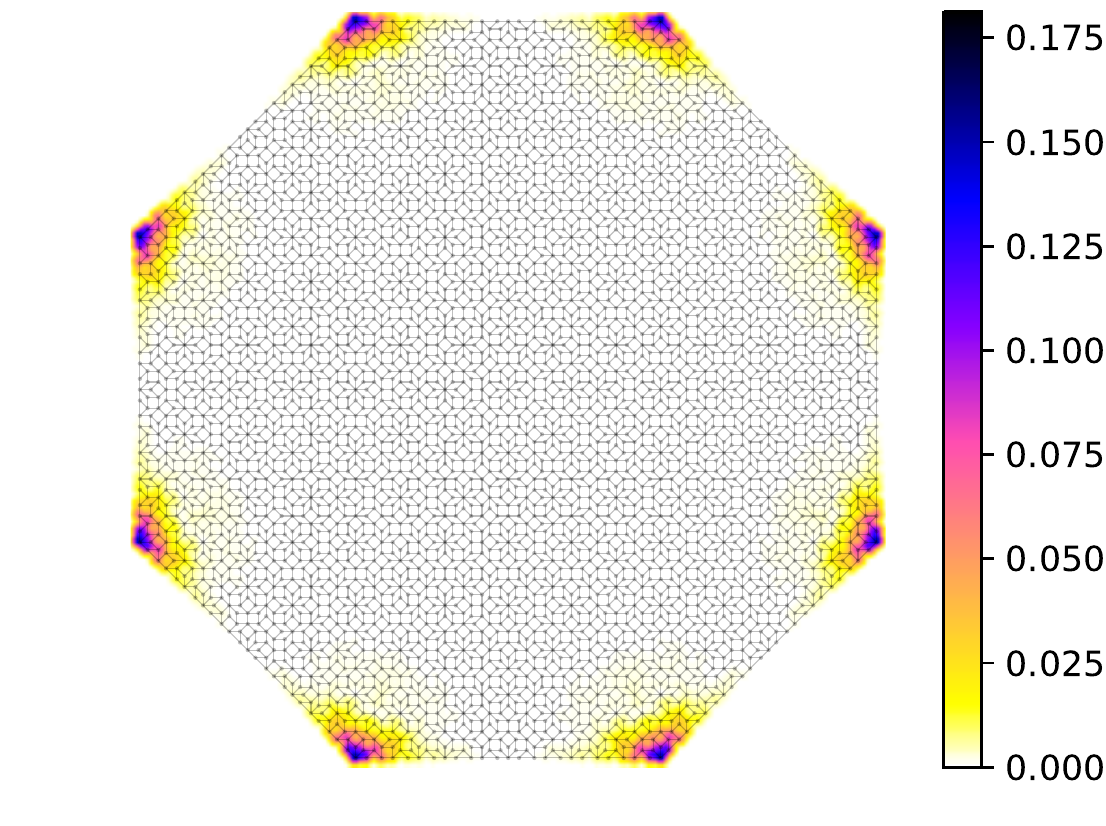}
    \includegraphics[width=0.5\columnwidth]{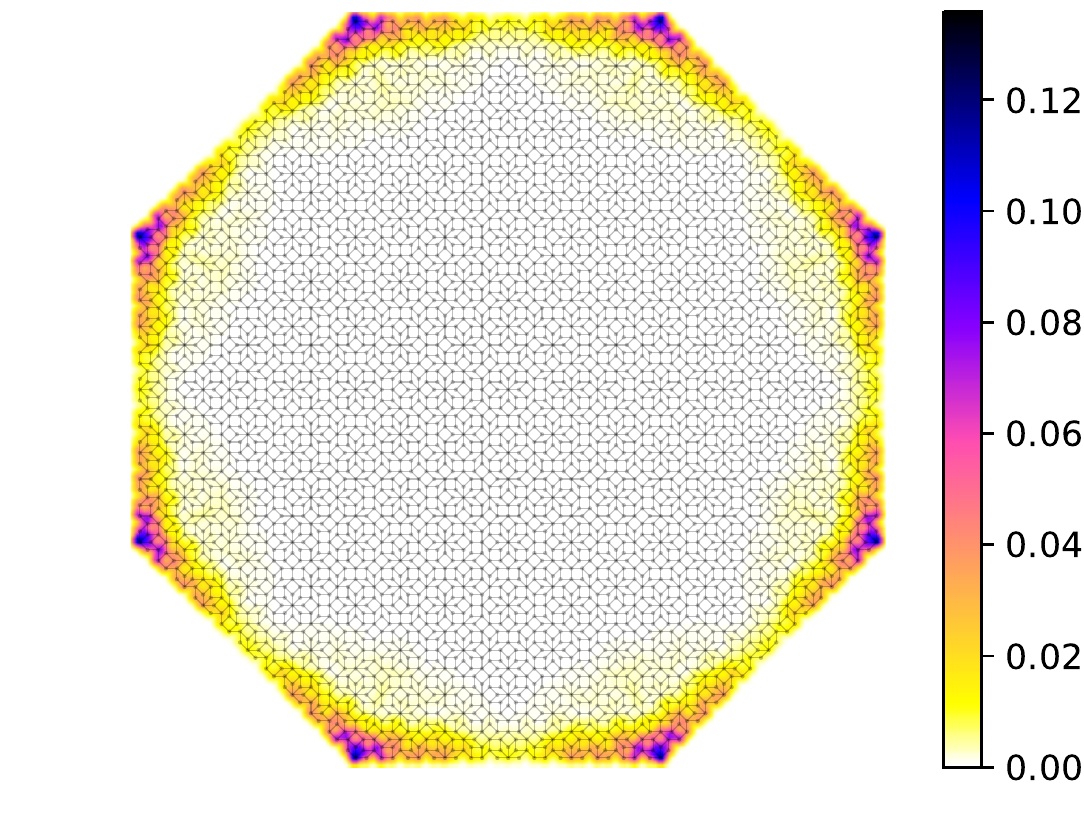}}
    \caption{Plot of real-space $|\psi|$ in the eight lowest energy states with (left) and without (right) edge triangles for $\Delta=t=V=1$ and $\mu=-1.5$.
    }
    \label{fig:prob_dist_terminations}
\end{figure}

In Figs.~1 and~2 of the main text we show finite samples of the quasicrystal that include extra hoppings forming triangles on the edge in order to make the edge straight.
For simplicity, these hoppings have the same form as the bulk hoppings in the same direction.
This makes the edge gap larger and the Majorana corner modes more localized.
As we argue in the main text, the topological protection of the Majorana corner modes is robust against deformations of the Hamiltonian on the edge, hence this choice is not essential for any of our findings.
To illustrate this, in Fig.~\ref{fig:prob_dist_terminations} we compare the Majorana wavefunctions with the two boundary conditions.

\begin{figure}[h]
    \includegraphics[width=0.5\columnwidth]{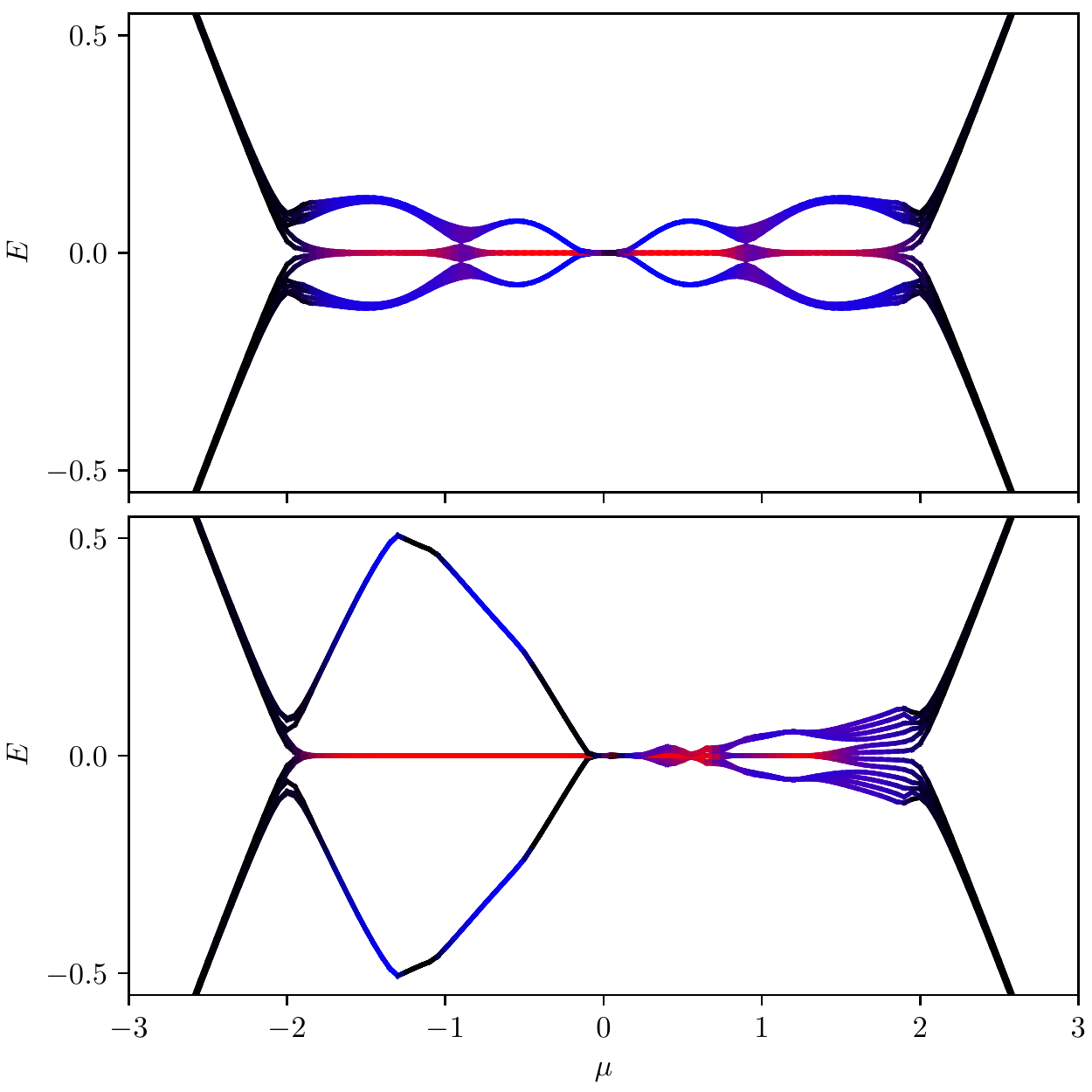}
    \caption{Spectrum of the 24 states closest to zero energy in a finite sample with different edge terminations: without edge triangles (top) and with edge triangles (bottom).
    The line color shows the weight of the state on the corners (red), edges (blue) and bulk (black).
    }
    \label{fig:edge_states_terminations}
\end{figure}

On the other hand, in Fig.~3 of the main text we use a sample without boundary triangles.
With this choice the edge spectrum is symmetric with respect to $\mu\to -\mu$.
As illustrated in Fig.~\ref{fig:edge_states_terminations}, including the edge hoppings makes the edge gap larger for $-2 < \mu < 0$, but smaller for $0 < \mu < 2$.
The bulk Hamiltonians are identical in the two cases, so the bulk density of states, the effective Hamiltonians and the resulting invariants are identical.
For all the plots we use a tiling constructed by applying the subdivision procedure $N=4$ times, which consists of $27137$ sites.
We define the width of the edge and corner regions as 5\% of the linear size of the system.

\section{Effective Hamiltonian and bulk invariant}

In our numerical results we approximate the effective Green's function
\begin{equation}
G_{\rm eff} (k)_{n,m}  = \left\langle k, n \right| G \left| k, m \right\rangle
\end{equation}
of the infinite system by a large finite sample using plane wave states are only nonzero in the interior, far from any edges or corners.
The Green's function of the full system is given by
\begin{equation}
G = \lim_{\eta\to0} \left(H + i \eta\right)^{-1}.
\end{equation}
Taking the thermodynamic limit of the system size going to infinity and the size of the interior region and $\eta$ fixed, the contribution of the edge states to $G_{\rm eff}$ decays exponentially.
Taking the $\eta\to 0$ limit after, we recover the bulk Green's function with all edge contributions removed.
The quasiperiodicity of the structure implies that there is a finite number of possible local patterns, and that these patches are repeated with equal density with every $C_8$-related orientation~\cite{baake2013aperiodic}.
Hence, in the thermodynamic limit, for a quasiperiodic locally generated Hamiltonian, $H_{\rm eff}$ is invariant under the onsite action of the symmetry even if the thermodynamic limit is taken by increasing samples that do not respect global 8-fold symmetry.
The effective Hamiltonian satisfies the symmetry constraints
\begin{eqnarray}
H_{\rm eff}(k) &=& U_{\cm} H_{\rm eff}\left(R_{\cm}^{-1} k \right) U_{\cm}^{-1}, \\
H_{\rm eff}(k) &=& - U_{\mathcal{P}} H_{\rm eff}\left(- k \right)^* U_{\mathcal{P}}^{-1}
\end{eqnarray}
where $U_{\cm}$ and $U_{\mathcal{P}}$ are the onsite unitary action of the $\cm$ and $\mathcal{P}$ operators and $R_{\cm}$ performs an 8-fold rotation on the momenta.

\section{Effect of other symmetries}

Throughout the manuscript we only use the $\cm$ symmetry of our model to establish topological protection. However, the full symmetry group is much larger.
Using the Qsymm software package~\cite{varjas2018qsymm}, we find that our quasicrystal HOTI model has a symmetry group with 64 elements.
It is generated by $\cm$, a mirror plane orthogonal to the plane of the system $M_x$, particle-hole symmetry $\mathcal{P}$, and an effective time reversal symmetry $\mathcal{T}$.
The full list and representations of the symmetry operators are available at~\cite{zenodo}.

\begin{figure}[tb]
    \includegraphics[width=\columnwidth]{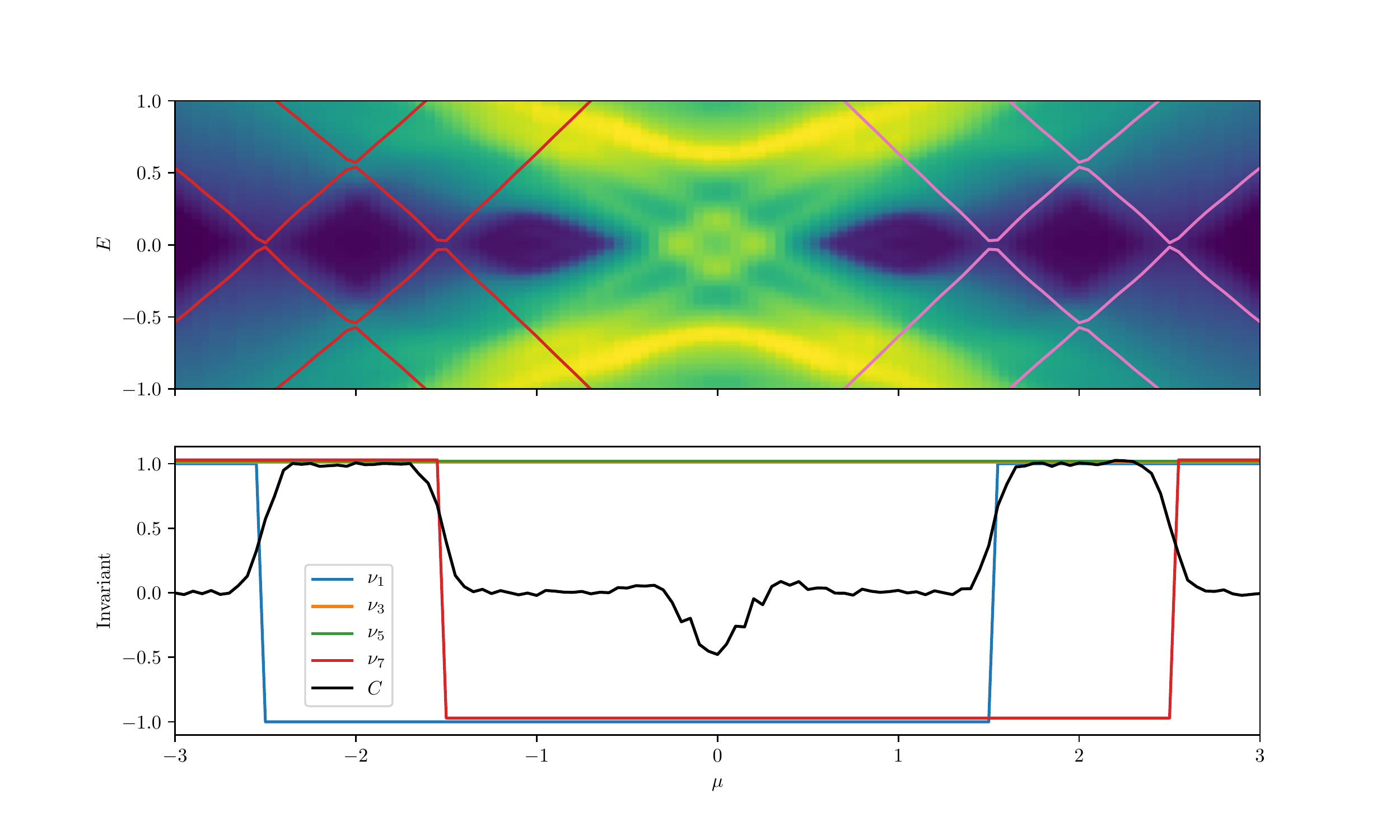}
    \caption{Top panel: Density of states in the bulk as a function energy and $\mu$ for $B=0.2$. Lighter colors denote larger densities. Overlaid is the spectrum of the effective Hamiltonian at $k=0$ (red) and $k=\Pi$ (pink). Bottom panel: Topological invariants as a function of $\mu$. The Chern number shows the intervening strong phase.}
    \label{fig:invariant_broken_T}
\end{figure}

Next, we investigate how the four stable $\mathbb{Z}_2$ invariants $\nu_n$ of $n \in [1, 3, 5, 7]$ defined in the main text relate to the other topological invariants.
First, we look at the strong $\mathbb{Z}$ invariant in class D.
As every other generalized $k$-point besides $0$ and $\Pi$ that is invariant under $\mathcal{P}$ ($k = -k$) has an even number of symmetry images under $C_8$, any band inversion there cannot change the strong class D $\mathbb{Z}$ Chern number invariant, whose parity is given by the product of the Pfaffians for all $\mathcal{P}$-invariant momenta.
Hence, the Chern number $C$ is constrained by $e^{i \pi C} = \prod_n \nu_n$.
As our model has an effective time reversal symmetry $\mathcal{T}$, the Chern number must vanish.
Adding a $\mathcal{T}$ breaking term allowed by $\cm$, $H_B = B \sigma_0 \tau_z$, we create an intervening phase with nonzero Chern number, hence gapless edges.
We illustrate this in Fig.~\ref{fig:invariant_broken_T}.
Restricting to phases with gapped bulk and edges, this puts a constraint on the $\mathbb{Z}_2$ invariants.

Another constraint comes from the classification with respect to the $C_4 = (C_8 M)^2$ symmetry.
The argument, analogous to the one with $C_8 M$, can be followed using $C_4$ instead, resulting in two $\mathbb{Z}_2$ indices $\xi_n$ for $n = [1, 3]$.
As long as $C_8M$ is a symmetry, the other $C_4$ invariant momenta have an even number of symmetry images and do not contribute to the invariant.
As the $C_4$ and $\mathcal{P}$ invariant blocks are the union of two $C_8$ blocks, the invariants are related as $\xi_1 = \nu_1 \nu_7$ and $\xi_3 = \nu_3 \nu_5$.
This leaves two independent, genuinely $C_8 M$ protected indices, $\nu_1$ and $\nu_3$, and we restrict to the case where the other invariants that do not rely on $C_8 M$ are trivial.

Our model has $\nu_1 = \nu_7 = -1$ and $\nu_3 = \nu_5 = +1$ in the topological phase.
We can construct a similar model with $\nu_1 = +1$ and $\nu_3 = -1$ and stack the two systems.
There is no obstruction to coupling the corners in a way that gaps out all Majoranas while keeping the global $C_8 M$ symmetry.
Hence, we find that the $\mathbb{Z}_2$ invariant characterizing the presence of corner Majoranas is the product $\nu_1 \nu_3$, while the corner modes do not distinguish between other phases in the richer $\mathbb{Z}_2^2$ bulk classification.

\section{Numerical methods}

We use the Kwant~\cite{Groth2014} software package to generate the tight-binding Hamiltonians with up to 27137 sites in the AB tiling obtained by $N=4$ subdivisions.
We find the lowest energy states using sparse diagonalization.

For the effective Hamiltonian, we use the Kernel Polynomial Method (KPM)~\cite{weisse2006kernel} to estimate the the Green's function of the full system.
KPM approximation of the Green's function using the Lorentz kernel is equivalent to replacing $1/x$ with a finite polynomial approximation of $1/(x + i\eta)$.
This only differs significantly from $1/x$ in a small region around 0, and $\eta$ depends on the full bandwidth and the number of moments used.
In the numerical calculations we use 200 moments, which results in $\eta = W / 200$ where $W$ is the full bandwidth (difference of the highest and lowest eigenenergy of the finite sample).
In the final step of inverting $G_{\rm eff}$ to get $H_{\rm eff}$, we take the Hermitian symmetrized $H_{\rm eff}$, which is equivalent to taking the real part of all eigenvalues of $H_{\rm eff}$, thus approximately recovering the $\eta\to 0$ limit.
For the numerical results shown here and in the main text we use an AB tiling with 27137 sites in total.
We calculate the approximate Green's function using a circular window in the center with diameter that is approximately 80\% of the system size, containing 16361 sites.

When evaluating $H_{\rm eff} (\Pi)$, we use the fact that $H_{\rm eff} (\Pi) = H'_{\rm eff} (0)$ with $H'$ the modified Hamiltonian where the sign of every nearest neighbor hopping is flipped.
We emphasize that this construction does not rely on the bipartite nature of the tiling we consider.
The invariants are robust against adding further neighbor hoppings, and the modified $H'$ is obtained by attaching a $-1$ factor to all hoppings between different sublattices and leaving the other hoppings invariant.

To calculate the Chern number we use the real space Chern marker formalism~\cite{Bianco2011} implemented using KPM~\cite{Varjas2019}.

Pfaffians are calculated using Pfapack~\cite{Wimmer2012}.

The data shown in the figures, as well as the code generating all of the data is available at~\cite{zenodo}.

\end{document}